\documentclass[conference]{IEEEtran}
\IEEEoverridecommandlockouts
\usepackage{cite}
\usepackage{amsmath,amssymb,amsfonts}
\usepackage{algorithmic}
\usepackage{graphicx}
\usepackage{textcomp}
\usepackage{xcolor}
\def\BibTeX{{\rm B\kern-.05em{\sc i\kern-.025em b}\kern-.08em
    T\kern-.1667em\lower.7ex\hbox{E}\kern-.125emX}}
\begin{document}

\title{MF-Hovernet: An Extension of Hovernet for Colon Nuclei Identification and Counting (CoNiC) Challenge\\
\thanks{
}
}

\author{\IEEEauthorblockN{ Vi Thi-Tuong Vo}
\IEEEauthorblockA{\textit{Department of AI Convergence} \\
\textit{Chonnam National University}\\
Gwangju, South of Korea \\
vothituongvi.cnu@gmail.com}
\and 
\IEEEauthorblockN{ Soo-Hyung Kim}
\IEEEauthorblockA{\textit{Department of AI Convergence} \\
\textit{Chonnam National University}\\
Gwangju, South of Korea \\
shkim@jnu.ac.kr}
\and
\IEEEauthorblockN{ Taebum Lee}
\IEEEauthorblockA{\textit{Department of Pathology} \\
\textit{Chonnam National University Medical School}\\
Gwangju, South of Korea \\
follyman@daum.net}
\and

}

\maketitle

\begin{abstract}
Nuclei Identification and Counting is the most important morphological feature of cancers, especially in the colon. Many deep learning-based methods have been proposed to deal with this problem. In this work, we construct an extension of Hovernet for nuclei identification and counting to address the problem named MF-Hovernet. Our proposed model is the combination of multiple filer block to Hovernet architecture. The current result shows the efficiency of multiple filter block to improve the performance of the original Hovernet model.

\end{abstract}


\section{Introduction}

In computational pathology, nuclear segmentation, classification and quantification within Haematoxylin \& Eosin stained histology images enables the extraction of interpretable cell-based features that can be used in downstream explainable models.

The Colon Nuclei Identification and Counting (CoNIC) Challenge \cite{graham2021conic} requires to development of algorithms that perform segmentation, classification and counting of 6 different types of nuclei: epithelial, lymphocyte, plasma, eosinophil, neutrophil or connective tissue.

This challenge includes two tasks. The first task requires participants to simultaneously segment nuclei within the tissue and classify each nucleus into one of six above categories. The second task is the regression task. The output of this task is how many nuclei of each class are present in each input image.

\section{Methodology}

\subsection{MF-Hovernet}
A proposed method is adopted to solve the problem, which combines the multiple filter block and Hovernet model \cite{graham2019hover} named MF-Hovernet.

 The multi-filter block \cite{vo2021effects} is used to deal with the issue by increasing the filter size instead of iteratively alleviating the image size. The multi-filter block is a stack of 3 convolution layers with different kernel size: 1 × 1, 3 × 3 and 5 × 5. The first convolution kernel has a filter size of 1 × 1.  We apply this convolution to reduce the size of the input vector and extract local features. While small kernels extract small complex features, the large kernel extract simpler features. Therefore, the next convolutional layer was set to 3 × 3 convolution kerel and uses a down-sampling size of 2 to obtain the global features. The last convolutional layer has a kernel size of 5 × 5 and a down-sampling size of 2. Each filter learns different features.

Our algorithm consists of Hovernet \cite{graham2019hover} and Preact-ResNet50 \cite{he2016identity} backbone. In addition, we changed each convolution layer in the original Hovernet model by the multiple filter block.

\begin{figure}[h]
\centering
\includegraphics[width=1\linewidth]{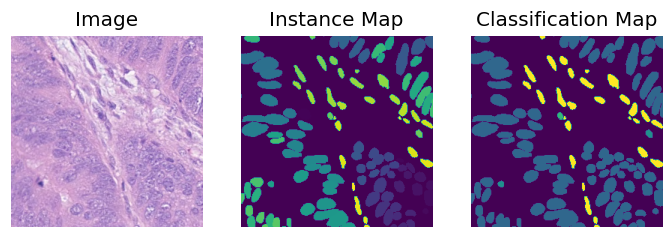}
\caption{Example of dataset. The dataset consists RGB images as raw image, instance map as ground truth for segmentation task, classification map as ground truth for classification task.}
\label{fig:input}
\end{figure}

\section{Experiments}

\subsection{Datasets}\label{AA}
We used the Lizard dataset \cite{graham2021lizard} of the CoNiC challenge, which is the current largest known publicly available dataset for instance segmentation in Computational Pathology. The Lizard dataset consists of Haematoxylin and Eosin stained histology images at 20x objective magnification (~0.5 microns/pixel) from 6 different data sources. For each image, an instance segmentation and a classification mask is provided . Within the dataset, each nucleus is assigned to one of the following categories: Epithelial, Lymphocyte, Plasma , Eosinophil, Neutrophil, Connective tissue.

\begin{figure}[h]
\centering
\includegraphics[width=0.8\linewidth]{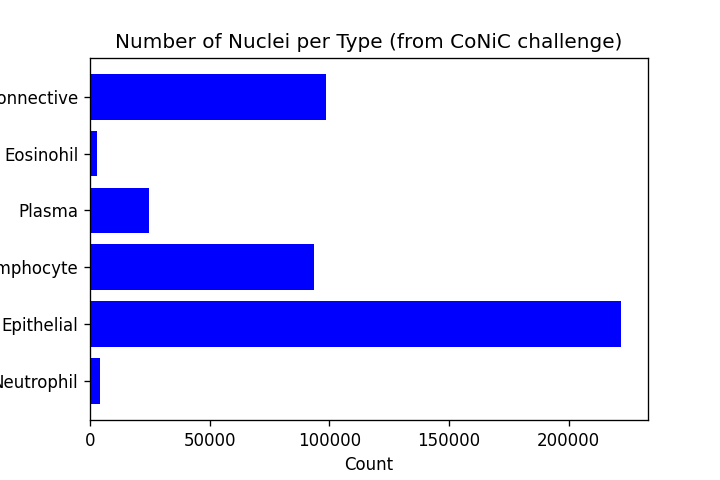}
\caption{The number of nuclei per type from CoNIC challenge. This dataset has the imbalance problem. The Epithelial cell has the most quantity. Meanwhile, the Eosinophil is the lowest. }
\label{fig:count}
\end{figure}

The dataset contains 4,981 non-overlapping patch images of size 256x256 provided in the following format: RGB images, Segmentation \& classification maps, Nuclei counts. The example of an input was shown in figure \ref{fig:input}.

The raw RGBimages and ground truth (segmentation maps and classification maps) are each stored as a single numpy array. The RGB image array has dimensions 4981x256x256x3, whereas the segmentation \& classification map array has dimensions 4981x256x256x2. Here, the instance segmentation map is the first channel and the classification map indicates in the second channel. Thre is a single csv file 
is provided for the nuclei counts, where each row corresponds to a given patch and the columns determine the counts for each type of nucleus. The row ordering is in line with the order of patches within the numpy files \cite{graham2021lizard}. Figure \ref{fig:count} shows the number of nuclei per type from CoNIC challenge. This dataset has the imbalance problem. The Epithelial cell has the most quantity with only 222017 cells. Meanwhile, the Eosinophil is the lowest with 2999 cells.

\subsection{Evaluation}
The multi-class panoptic quality (mPQ) in \eqref{mpq} is used to determine the performance of nuclear instance segmentation and classification. The PQ \cite{kirillov2019panoptic} of each type t  is defined as \eqref{pqt}.

\begin{equation}
mPQ = \frac{1}{T} \sum_{1} PQ_t     \label{mpq}
\end{equation}
where 
\begin{equation}
PQ_t = \frac{|TP_t|}{TP_t + \frac{1}{2} |FP_t| + \frac{1}{2} |FN_t|} \times \frac{\sum_{(x_t,y_t)\in TP} IOU (x_t, y_t)}{|TP_t|}      \label{pqt}
\end{equation}

The $R^2_t$ is used for regression task as shown in \eqref{r2}.

\begin{equation}
R^2_t = 1 - \frac{RSS_t}{TSS_t}    \label{r2}
\end{equation}
Here, $RSS$ and $TSS$ stands for the sum of squares of residuals and the total sum of squares, respectively.

\subsection{Experiment setup}

 We split our training data into 80\% for training and 20\% for validation with 5-fold cross-validation. For network training, we used a patch as input with size of 256 × 256 pixels and a batch size of 6. Furthermore, no data augmentation technique was applied to this experiment. Our method is the end-to-end model. We trained the network with a learning rate of $10^4$ for 100 epochs until convergence. For loss computation, we calculated the multiple term regression loss such as mean squared error loss, cross-entropy loss \cite{zhang2018generalized}, dice loss \cite{li2019dice} for simultaneous nuclei segmentation and classification domain which was presented in \cite{graham2019hover}. Model selection was guided by the highest performance on the validation set.  The Adam optimizer \cite{kingma2014adam} is used as the optimization method for model training. All models are implemented using the PyTorch framework \cite{paszke2017automatic}.

\subsection{Experiment result}
\begin{figure}[h]
\centering
\includegraphics[width=1\linewidth]{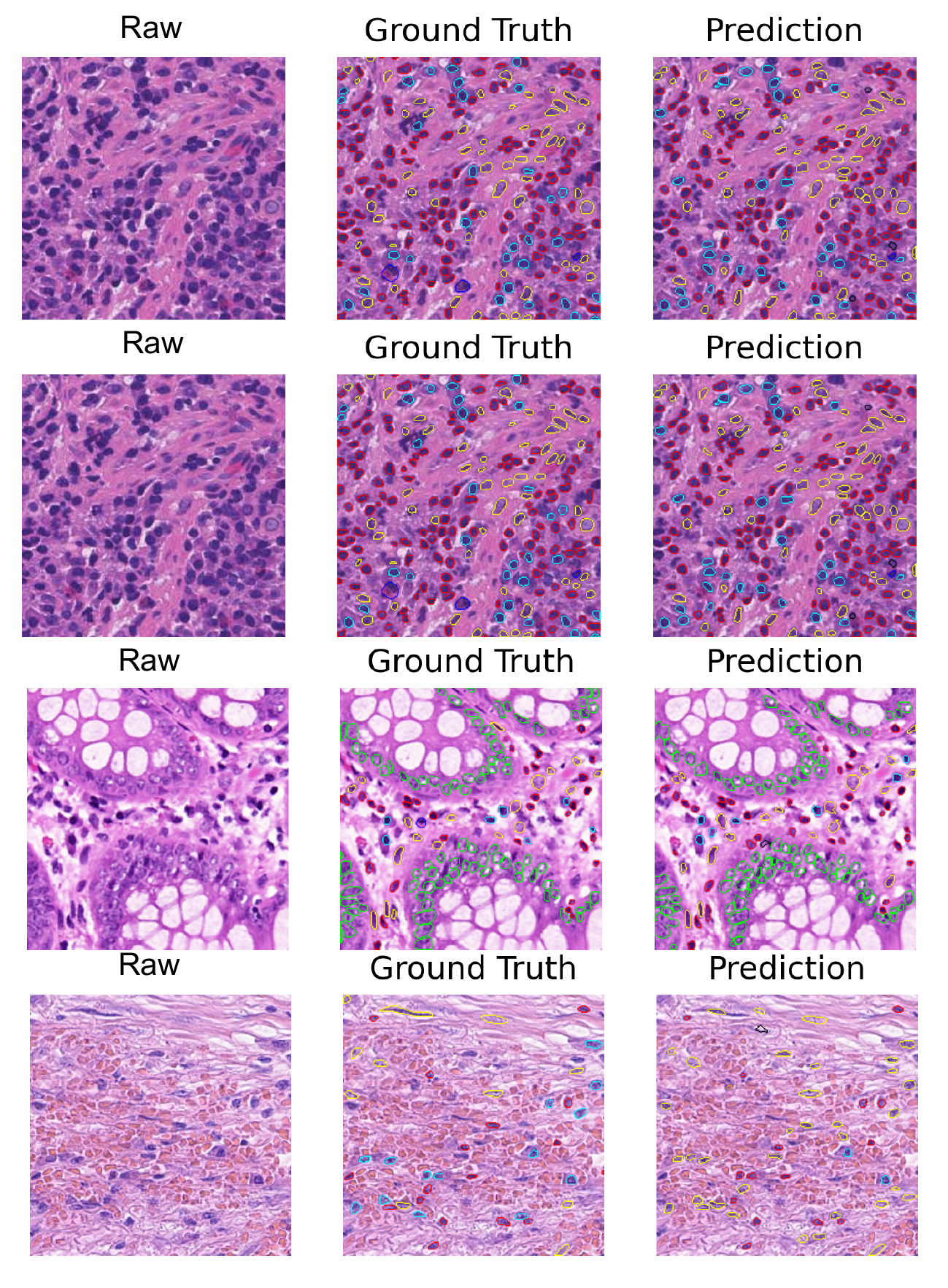}
\caption{Example of nuclei identification which predicted by MF-Hovernet. The first column is the raw images. The second column presents the ground truth provided by the challenge. The last column is the predicted images from MF-Hovernet. }
\label{fig:cellresult}
\end{figure}

\begin{figure}[h]
\centering
\includegraphics[width=1\linewidth]{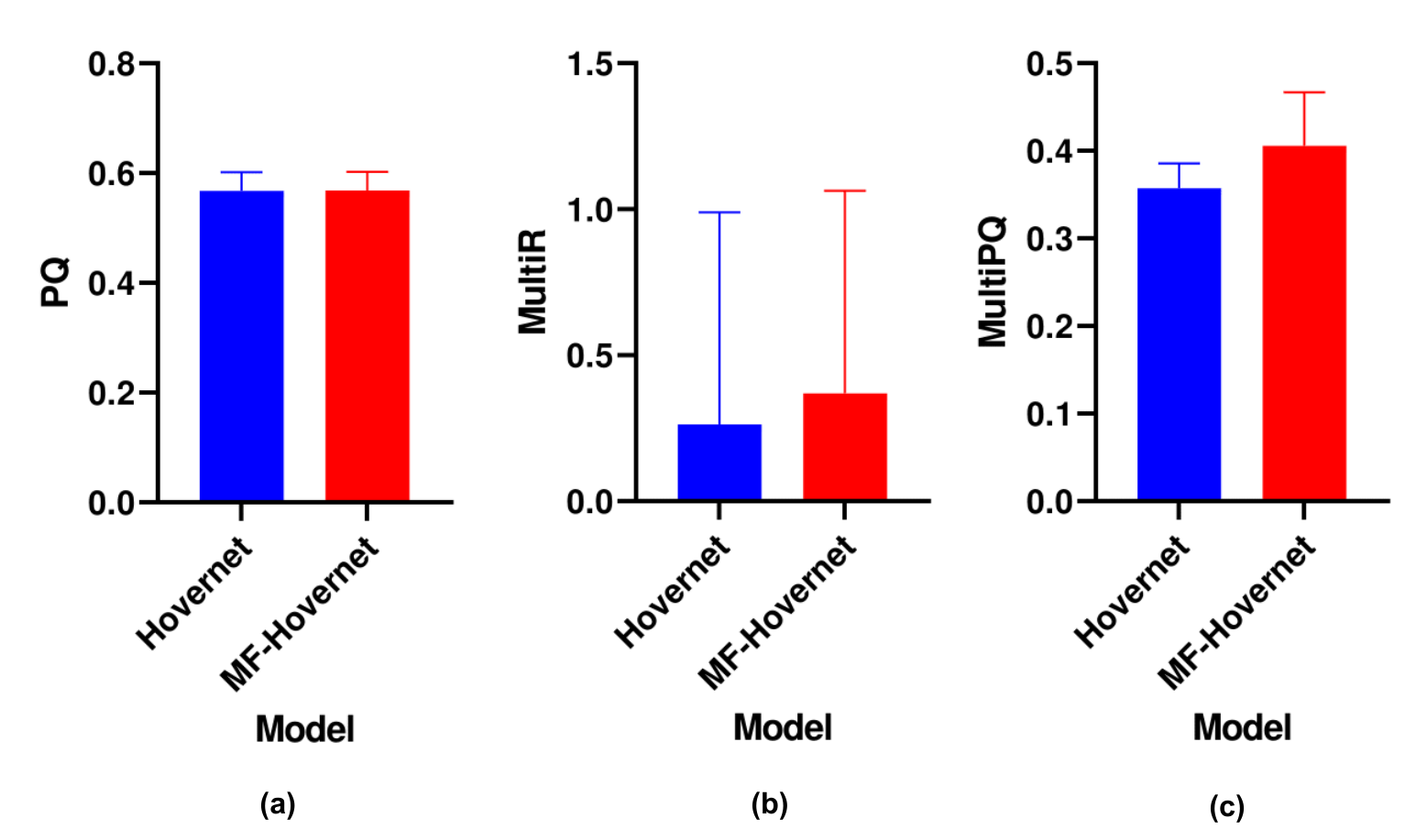}
\caption{Result comparison between original Hovernet and MF-Hovernet via 3 measurements: PQ, MultiR, MultiPQ based on 5-folds.}
\label{fig:result}
\end{figure}

Figure \ref{fig:result} shows the result comparison between original Hovernet and MF-Hovernet via 3 measurements: PQ, MultiR, MultiPQ on the validation set. The MF-Hovernet achieved higher results than the original Hovernet in 2 metrics: MultiR and MultiPQ.

This means the multiple filter block help to improve the performance of the original Hovernet. The current results show the efficiency of multiple filter block to enhance the performance of the original Hovernet model.

Figure \ref{fig:cellresult} presents some example results which predicted from our model. The first column is the raw images. The second column presents the ground truth provided by the challenge. The last column is the predicted images from MF-Hovernet. There are four samples was shown in figure \ref{fig:cellresult} corresponding with four rows.

\section{Conclusions}
This paper reported a method for nuclei identification and counting from pathology images in the CoNiC challenge. Our method integrates the multiple filter block to Hovernet for segmenting and counting nuclei on pathology images. The current result shows the efficiency of multiple filter block to improve the performance of the original Hovernet model. In the future, we aim to improve our method and investigate the performance of the test set of the CoNiC challenge.

\bibliographystyle{IEEEtran} 
\bibliography{refs}

\end{document}